\documentclass{article}

\PassOptionsToPackage{numbers, compress}{natbib}

\usepackage[final]{neurips_2024}

\usepackage{graphicx}
\usepackage[utf8]{inputenc} 
\usepackage[T1]{fontenc}    
\usepackage{hyperref}       
\usepackage{url}            
\usepackage{booktabs}       
\usepackage{amsfonts}       
\usepackage{nicefrac}       
\usepackage{microtype}      
\usepackage{xcolor}         
\usepackage{svg}
\usepackage{amsmath}
\usepackage{multirow}
\usepackage{array}
\usepackage{geometry}
\usepackage{colortbl}
\usepackage{quoting}
\usepackage{subcaption}
\quotingsetup{leftmargin=1cm}
\usepackage{enumitem}
\usepackage{bbm}

\newcommand{\tab}[1]{Table~\ref{#1}}
\newcommand{\fig}[1]{Fig.~\ref{#1}}

\newcommand{\ie}{\textit{i.e.}, }

\newcommand{\etal}{\textit{et al.}, }

\newcommand{\none}{\texttt{None}}
\newcommand{\gpto}{\texttt{gpt-4o}}
\newcommand{\gptm}{\texttt{gpt-4o-mini}}
\newcommand{\llama}{\texttt{Llama}}
\newcommand{\bert}{\texttt{BERT}}
\newcommand{\dbert}{\texttt{DistilBERT}}

\title{Towards Optimizing SQL Generation\\ via LLM Routing}

\author{%
Mohammadhossein Malekpour \quad Nour Shaheen \quad Foutse Khomh \quad Amine Mhedhbi\\
Polytechnique Montr\'eal\\
\texttt{\{mohammadhossein.malekpour,nour.shaheen,foutse.khomh,amine.mhedhbi\}}\\\texttt{@polymtl.ca}\\
}

\begin{document}

\maketitle

\begin{abstract}
Text-to-SQL enables users to interact with databases through natural language, simplifying access to structured data. Although highly capable large language models (LLMs) achieve strong accuracy for complex queries, they incur unnecessary latency and dollar cost for simpler ones. In this paper, we introduce the first LLM routing approach for Text-to-SQL, which dynamically selects the most cost-effective LLM capable of generating accurate SQL for each query.

We present two routing strategies (score- and classification-based) that achieve accuracy comparable to the most capable LLM while reducing costs. We design the routers for ease of training and efficient inference. In our experiments, we highlight a practical and explainable accuracy-cost trade-off on the BIRD dataset.
\end{abstract}

\vspace{-0.6em}

\section{Introduction}
\label{sec:introduction}
\vspace{-0.7em}

In recent years, Text-to-SQL has gained significant momentum with deployments within enterprise solutions to transform data accessibility~\cite{androutsopoulos1995naturallanguageinterfacesdatabases,DBLP:journals/ftdb/QuamarELO22}. 
Text-to-SQL democratizes access to structured data, allowing non-experts to interact with databases directly without requiring data engineering expertise. For SQL analysts, it enhances their workflows by supporting query authoring, dataset exploration, and report generation. A major use case lies in iterative data exploration, where the \emph{complexity of user queries can range widely—from simple row retrievals to multi-way joins with aggregations}.

Current state-of-the-art Text-to-SQL approaches follow a multi-stage pipeline \cite{hong2024nextgenerationdatabaseinterfacessurvey,Li2024TheDO, liu2024surveynl2sqllargelanguage,zhang2024naturallanguageinterfacestabular}. The pipeline consists of two main phases: (i) retrieval of contextual information--such as schema elements, examples, and instructions relevant to the query--and (ii) SQL generation. 
Current enterprise solutions use highly capable LLMs for SQL generation to handle highly complex queries~\cite{maamari2024deathschemalinkingtexttosql,maamari2024endtoendtexttosqlgenerationanalytics}. 
While this is essential for complex queries, such highly capable LLMs introduce considerable latency and incur higher costs for simpler ones, such as inspecting a few rows from a table. 
This in turn, can negatively impact both user experience and the average cost per query.

Leading Text-to-SQL benchmarks such as {BIRD}~\cite{li2023llmservedatabaseinterface} rank submissions only based on accuracy while also indicating the model used on the leaderboard. 
Benchmarks typically assume a single-model approach, which mirrors the same inefficiencies found in enterprise deployment—namely, using the most capable models for all queries regardless of complexity. For instance, six of the top ten solutions on BIRD use \texttt{GPT-4o} or \texttt{Gemini}. 

\begin{figure}[t!]
  \centering
  \includegraphics[width=1.\columnwidth]{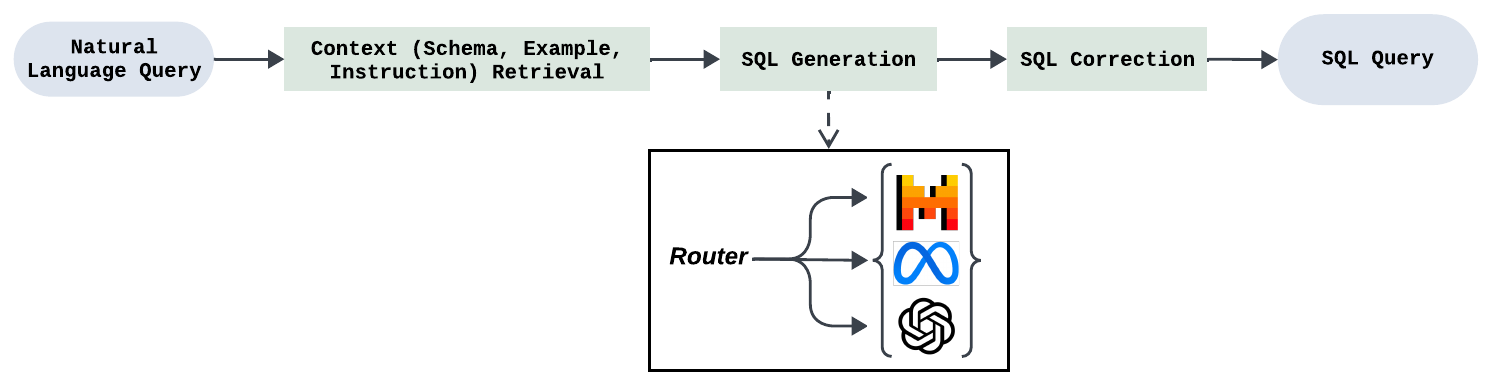}
  \caption{A multi-stage Text-to-SQL pipeline with an LLM router within the generation stage.}
  \vspace{-1em}
  \label{fig:architecture}
\end{figure}

To handle efficiently the varying levels of complexity, 
\emph{we investigate the implementation of LLM routers for Text-to-SQL}. They route a query to the weakest, yet cheaper and faster model, capable of generating accurate SQL. 
We propose two routing approaches that achieve an accuracy close to that of the most capable LLM, always outperforming the second-best, and reducing costs by up to $1.4\times$. This cost reduction is substantial for enterprise deployments of analytics or SQL assistants with a high volume of queries. 
These routers are designed to be easy to train and efficient at inference. \fig{fig:architecture} depicts our pipelines with an SQL generation router. 

\section{Preliminaries}
\label{sec:related-work}

\subsection{Problem Formulation}

Given a set of $N$ models \{ $\mathcal{M}_0$, $\mathcal{M}_1$, $\dots$, $\mathcal{M}_{N-1}$ \} and a natural language query $Q$, our objective is to identify the \emph{weakest} model $\mathcal{M}_i$ that can accurately generate corresponding SQL for $Q$. 
Here, we define the weakest model as the one with the lowest SQL generation capability. 
In practice, this model typically exhibits the lowest latency and incurs the least dollar cost. 
We assume a total ordering of model strength, such that $\mathcal{M}_i$ is considered weaker than $\mathcal{M}_j$ if $i < j$. Therefore, we formulate the problem as an optimization task: select the smallest $l$ such that $\mathcal{M}_l$ produces accurate SQL for $Q$.

Furthermore, we assume access to a dataset $H$, 
consisting of prior natural language (NL) queries, predicted SQL outputs, and corresponding ground-truth SQL for each of the $N$ models. 
In practice, these examples can be collected from execution logs, user feedback, and manual labeling.

Accordingly, our goal is to learn an $N$-ary routing function that maps a query $Q$ to the minimal label $l \in [0, N]$. 
Here, a label $l \in [0, N-1]$ indicates that $\mathcal{M}_l$ is the weakest model capable of generating accurate SQL for $Q$. 
A prediction of $N$ denotes that no model within the set is capable of doing so.

\subsection{Related Work}

LLM Routing has been studied as a binary routing problem \cite{DBLP:journals/corr/abs-2404-14618,DBLP:journals/corr/abs-2406-18665} for tasks such as question answering, summarization, and information extraction. 
To our knowledge, this is the first study of LLM $N$-ary routing and the first for the Text-to-SQL task.

Other prior work uses an ensemble of models for generation, as proposed by Jiang \etal~\cite{DBLP:journals/corr/abs-2306-02561}. 
The approaches use multiple models to generate a set of answers instead of generating a single one and then use pairwise comparison and fusion of top-candidate to generate a final answer. 
While this approach is promising to maximize accuracy, it goes against the objective of cost minimization. 
A cascade approach has also been previously employed \cite{DBLP:journals/corr/abs-2305-05176}. However, it cannot be used for Text-to-SQL. 
Starting with the weakest model, it is infeasible to determine whether the generated SQL is accurate or meets a quality threshold 
before falling back to the next available stronger model. 

\subsection{Metrics}

To evaluate the effectiveness of generated SQL, we use \emph{execution accuracy} ($EX$) as the primary metric ~\cite{li2023llmservedatabaseinterface,DBLP:journals/corr/abs-1809-08887}. 
Specifically, $EX(S)$ denotes the proportion of queries in the evaluation set $S$ for which the output relation (from the predicted SQL) matches the ground-truth relation, where $0$~$\leq$~$EX(S)$~$\leq$~$1$.
Here, matching relations are defined as those containing identical tuples, independent of attribute ordering. 
Our objective is to maximize $EX(S)$ while concurrently choosing the weakest model. 
In this work, we consider dollar cost as the cost we aim to minimize. 

\section{Methods}
\label{sec:method}

We consider two approaches. 
The first is a regression approach where we assign a score per model predicting the capability of generating accurate SQL for $Q$. 
We say a model is capable of generating accurate SQL if its score is above an input threshold. 
For all models above that threshold, we pick the weakest. 
The second is a classification approach where a router model predicts $l \in [0,N]$ as defined in the problem formulation.

If all scores of the regression approach are under the threshold or the classification approach predicts $N$, then we are predicting that no model is capable of generating accurate SQL. 
As such, dependent on the use case, we can decide to not attempt to generate SQL or we can route to one of the $N$ models. 

\subsection{Score-based Routing}

A regression-based router predicts for each model $\mathcal{M}_i$, a score $P(EX=1_{\mathcal{M}_i}|Q)$, \ie the probability of generating accurate SQL for an input query $Q$. 
The router has the input parameter $\alpha$: the score threshold such that if $P(EX=1_{_i}|Q) < \alpha$ then $\mathcal{M}_i$ is said to be incapable of generating the query. 
As such, this router minimizes $l$ such that $P(EX=1_{\mathcal{M}_l}|Q) \geq \alpha$. 

If the scores indicate that no model is capable of generating accurate SQL for $Q$, we can choose to not make an attempt or to route to one of the models for a different accuracy-cost trade-off. 
To maximize accuracy in our implementation, we choose to route to the strongest model. 

We implement the scoring function using $H$ 
for each of the $N$ models. 
Given an input query $Q$, we first find $\mathcal{Q}_k$: the top-$K$ similar queries in $H$. 
For each model $\mathcal{M}$, 
we set $P(EX=1_{\mathcal{M}_i}|Q) = EX{(\mathcal{Q}_k)_{\mathcal{M}_i}}$. 
We then filter the models ($EX{(\mathcal{Q}_k)_{\mathcal{M}_i}} \geq \alpha$) and pick the weakest. 
We denote this router based on its two parameters as $R_{\alpha}^{K}$.

\subsection{Classification-based Routing}

We use a distilled \bert-style encoder model (\dbert~\cite{DBLP:journals/corr/abs-1910-01108}), 
which we train on $H$ to learn the function $R_{BERT}$. 
Given an input query $Q$, we input the NL query and relevant retrieved schema to predict the label $l$ indicating the weakest model $\mathcal{M}_l$ predicting $EX(Q)$ $=$ $1_{\mathcal{M}_i}$. 
If $R_{BERT}$ predicts $N$ then it is predicting that none of the models can generate accurate SQL. Similarly to the other router, to maximize accuracy in our implementation, we choose to route to the strongest model. 

\section{Experiments}
\label{sec:method}

\subsection{Datasets}

We conducted our experiments using the BIRD dataset~\cite{li2023llmservedatabaseinterface},
which is widely considered to be the most challenging Text-to-SQL benchmark. 
Our evaluation set consisted of all $1534$ queries in the dev dataset and
our training set, acting as $H$, consisted of $9428$ queries. 

\subsection{Models}
\label{subsec:models}

\textbf{Specific models and strength.} We used three different LLMs for SQL generation: i) \texttt{GPT-4o}; ii) \texttt{GPT-4o-mini}; and iii) Llama: \texttt{Llama-3.1-8B-instruct-q4\_0}. 
To order their SQL generation capability, we used a simplified Text-to-SQL pipeline consisting of a single attempt at generation while adding the whole schema to the LLM context. 
We evaluate the pipeline by selecting 10\% of the queries of each database in BIRD uniformly at random. 
We find \llama~as the weakest model with EX $0.34$, then \gptm~with EX $0.48$, and \gpto~as the strongest with EX $0.55$. 

\textbf{Cost.} 
In our experiments, we only analyzed dollar cost and forgo latency due to setup challenges. 
We use \llama~as a local model and hence associate no dollar cost with it and use OpenAI services to access \gpto~and \gptm. Instead of using current token prices of OpenAI, we use a normalized unit cost where we set \gptm~input and output token cost to $1$ and \gpto to $16.6\times$ (multiplicative price difference).\footnote{Numbers obtained based on OpenAI's offering as of 1 Oct. 2024}
We also used OpenAI's \texttt{text-embedding-3-small} model with cosine similarity to find the top $K$ similar queries in $H$ for the score-based router. 
The embedding cost is negligible per input NL query when compared with token usage and is therefore ignored. 

\textbf{Score-based Router Implementation.} 
For a score-based router ($R_{k}^{\alpha}$), we have to select the two parameters $\alpha$ and $K$. $\alpha$ (threshold score) is the minimum proportion out of the $K$ similar queries for which a model had to generate accurate SQL to be considerate a candidate model for generation. 
We run a grid search over 10\% of train queries as done before and for each input query, we remove the queries in the same database from being chosen as similar ones. 
We pick $\alpha > 0.5$ as it means intuitively some level of robustness where more than half of the $K$ similar queries were generated successfully. 
For each $\alpha \in \{ 0.6, 0.7, 0.8, 0.9 \}$, we do a search over $k \in [5,50]$. 
We find that setting $\alpha$ to $0.9$ is overly restrictive indicating $N$ for every input query, 
\ie no model is capable of generate accurate SQL. 
As such, we only consider $\{ 0.6, 0.7, 0.8 \}$. 
We choose the $K$ that maximizes $EX$ while having routed to each model at least once. 
We end up with three models with a difference less than $0.5\%$ EX between them: $R_{24}^{0.6}$, $R_{25}^{0.7}$, and $R_{10}^{0.8}$.

\textbf{Classification-based Router Implementation.} 
We fine-tuned \dbert~\cite{DBLP:journals/corr/abs-1910-01108} on the fully labeled train set $H$. 
We removed from $H$, all queries in the retail\_world database due to schema missing and all duplicate user questions. 
We split $H$ 80-20 on db\_id first (ensuring databases in train are not in validation and vis-versa). This led to a train set of $7346$ queries and a validation set of $1697$. 
When training on $H$ and evaluating on the dev set, to retrieved the relevant schema using the TCSL schema linking approach~\cite{talaei2024chess,maamari2024deathschemalinkingtexttosql}.
We trained for $4$ epochs, with a batch size of $32$, and a learning rate of $1.00E$$-$$04$. We sampled each batch using a weighted random sampler, the weights correspond to the inverse of the frequency of each label within to the training data. 

\subsection{Results}
\label{sec:results}

\renewcommand{\arraystretch}{1.4}
\begin{table}[t!]
    \centering
    \caption{EX\%, model distribution, cost compared to \gpto~for each generation approach on the BIRD dev set, \ie three routers from above and the use of the three base models: (i) \gpto; (ii) \gptm; and (iii) \llama: \texttt{llama3.1:8b-instruct-q4\_0}.}\vspace{0.4em}
    \begin{tabular}{lcrrrrr}
        \rowcolor[HTML]{F1F1F7}\hline
        & 
        & \multicolumn{4}{c}{\textbf{\# Queries}}
        & \\\cline{3-6}
        \rowcolor[HTML]{F1F1F7}
        \multirow{-2}{*}{\textbf{Gen.}} & 
        \multirow{-2}{*}{\textbf{EX\%}} & 
        \gpto & \gptm & \llama & \none &
        \multirow{-2}{*}{\textbf{\$ Reduc.}} \\\hline
        \texttt{4o} & $61.02$ & $1534$ ($100$\%) & - & - & - & \textbf{1x} \\
        \rowcolor[gray]{0.97} 
        \texttt{4o-mini} & $49.22$ & - & $1534$ ($100$\%) & - & - & \textbf{16.6x}\\
        \llama & $29.34$ & - & - & $1534$ ($100$\%) & - & $\infty$ \\\hline
        \rowcolor[gray]{0.97} $R_{25}^{0.7}$ & $60.14$ & $197$ ($13$\%) & $88$ ($6$\%) & $5$ ($0$\%) & $1243$ ($81$\%) & \textbf{1.1x} \\
        $R_{10}^{0.8}$ & $59.42$ & $160$ ($10$\%) & $127$ ($8$\%) & $26$ ($2$\%) & $1220$ ($80$\%) & \textbf{1.1x} \\
        \rowcolor[gray]{0.97} $R_{24}^{0.6}$ & $57.92$ & $324$ ($21$\%) & $265$ ($17$\%) & $54$ ($4$\%) & $890$ ($58$\%) & \textbf{1.3x} \\
        $R_{BERT}$ & $55.21$ & $118$ ($8$\%) & $311$ ($20$\%) & $167$ ($5$\%) & $938$ ($61$\%) & \textbf{1.4x} \\
        \hline
    \end{tabular}
    \label{tab:ex}
    \vspace{-1em}
\end{table}

We evaluate our routers $R_{24}^{0.6}$, $R_{25}^{0.7}$, $R_{10}^{0.8}$, and $R_{BERT}$ on all $1534$ dev queries. 
Recall that all models, route to the strongest model (\gpto) in case $N$ is predicted. 
By analyzing the successful and failed query sets on the three base models (Details in Appendix~\ref{sec:appendix_failure_analysis}), 
we find that each weaker model has a large subset of failed and successful queries with a stronger one. 
As such, we expect routing to lower the cost while at best keeping the best model's EX. 

\tab{tab:ex} summarizes the EX, model routing distribution and relative cost to the strongest model (\gpto) for each generation approach on BIRD's dev set. 
Our routers are up to 1.4x cheaper while being close in EX. 
With both routers, we lose some accuracy for a lower cost. 
The accuracy-cost trade-off on BIRD is easiest to explain through the parameters of the score-based router (Appendix~\ref{sec:k_alpha_effect}). 

In this ongoing work, we aim next to assess whether an NL query with relevant schema is enough to classify its complexity as done in $R_{BERT}$. 
Furthermore, we plan to explore routing across more datasets and within more stages such as correction and schema linking instead of just generation. 

\section{Conclusion}
\label{sec:conclusion}

In this work, 
we investigated two LLM routing approaches for text-to-SQL on the BIRD benchmark. 
We believe that cost-based optimization techniques are important for enterprise-level systems where not only accuracy but also cost are critical. 
In the future, we intend to explore a larger scope of empirical analysis with similar techniques to improve both the execution accuracy and cost. 

\clearpage

\bibliographystyle{abbrv}
\bibliography{references}

\clearpage

\appendix

\section{Analysis of Failure Cases}
\label{sec:appendix_failure_analysis}

\begin{figure}[h]
    \centering
    \begin{subfigure}[t]{0.48\textwidth}
        \centering
        \includegraphics[width=\textwidth]{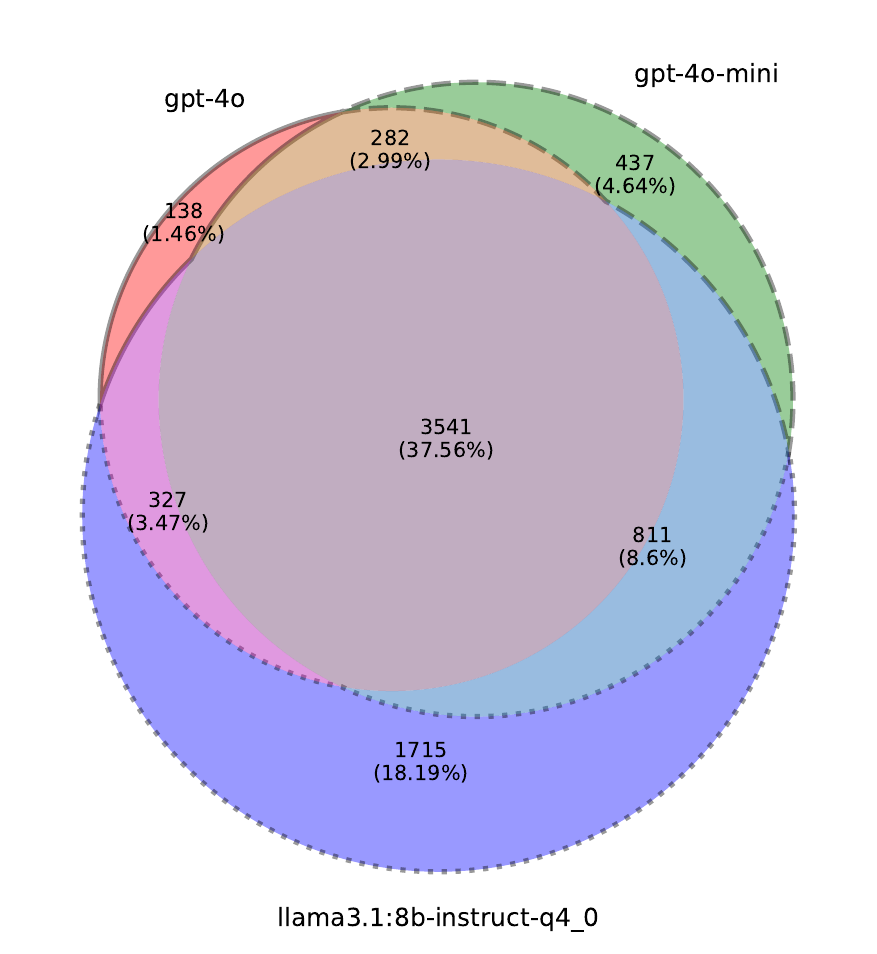}
        \caption{Venn Diagram of Failure Cases}
        \label{fig:failed_answers_venn}
    \end{subfigure}%
    \hfill
    \begin{subfigure}[t]{0.48\textwidth}
        \centering
        \includegraphics[width=\textwidth]{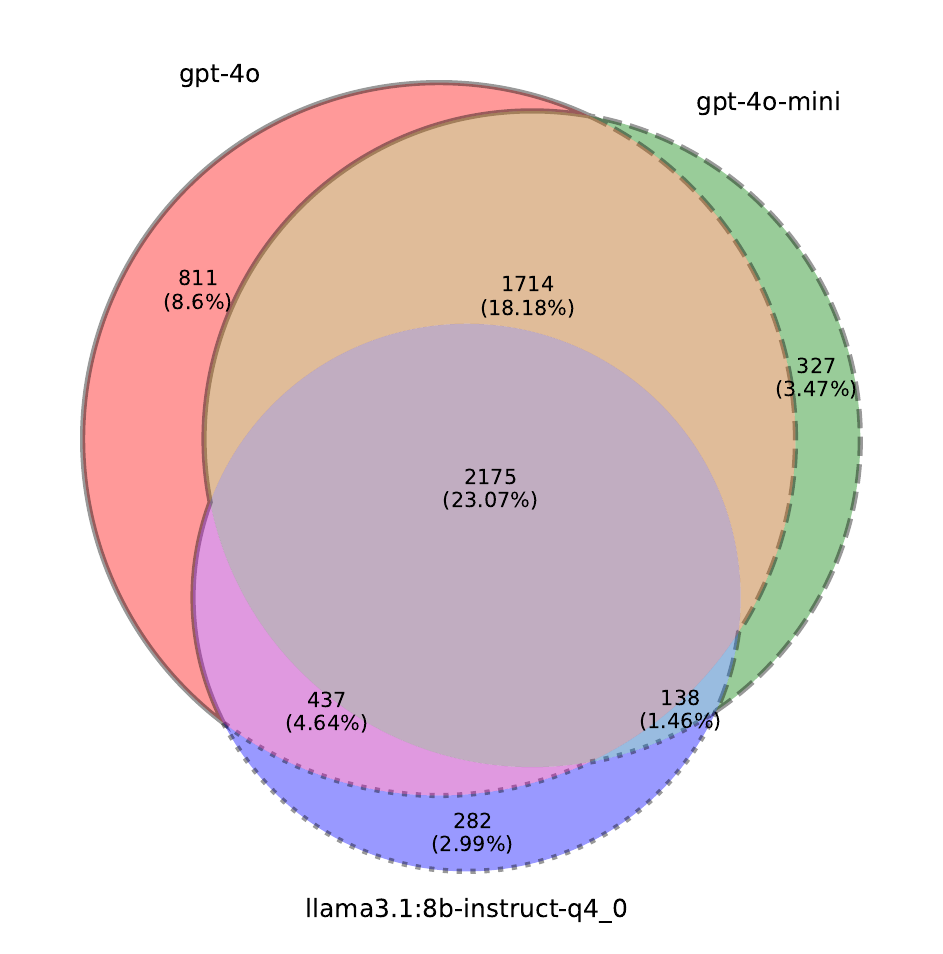}
        \caption{Venn Diagram of Correct Cases}
        \label{fig:correct_answers_venn}
    \end{subfigure}
    \caption{Distribution of Failure and Correct Cases across \texttt{gpt-4o}, \texttt{gpt-4o-mini}, and \texttt{llama3.1:8b-instruct-q4\_0}}
    \label{fig:venn_diagrams}
\end{figure}

We analyze the performance of each model in terms of correct and failed query sets cases, highlighting their differences in handling SQL generation. 
Figures \ref{fig:failed_answers_venn} and \ref{fig:correct_answers_venn} illustrate the distribution of failed and correct predictions across the three models: \texttt{gpt-4o}, \texttt{gpt-4o-mini}, and \texttt{llama3.1:8b-instruct-q4\_0}.

\begin{table}[h]
    \centering
    \caption{Failure Cases across Models}
    \begin{tabular}{lcc}
        \toprule
        \textbf{Failure Cases} & \textbf{Count} & \textbf{Percentage} \\
        \midrule
        \texttt{gpt-4o} & 4288 & 45.49\% \\
        \texttt{gpt-4o-mini} & 5071 & 53.79\% \\
        \texttt{llama3.1:8b-instruct-q4\_0} & 6394 & 67.83\% \\
        \midrule
        Intersection \texttt{gpt-4o} \& \texttt{gpt-4o-mini} \& \texttt{llama3.1:8b-instruct-q4\_0} & 3541 & 37.56\% \\
        Intersection \texttt{gpt-4o} \& \texttt{gpt-4o-mini} & 3823 & 40.55\% \\
        Intersection \texttt{gpt-4o} \& \texttt{llama3.1:8b-instruct-q4\_0} & 3868 & 41.03\% \\
        Intersection \texttt{gpt-4o-mini} \& \texttt{llama3.1:8b-instruct-q4\_0} & 4352 & 46.17\% \\
        \bottomrule
    \end{tabular}
\end{table}


The difference in EX shows a clear gap in capability between models. 
We find that 37.56\% of the queries failed across all three models with a high common failed percentage of queries between every pair from 45.49\%--67.83\%. 
This indicates that when routing, it is unlikely to improve EX beyond that of the strongest model and that we expect a cost decrease only.


\section{Effect of Varying $K$ and $\alpha$ on Execution Accuracy and Model Distribution}
\label{sec:k_alpha_effect}

We analyze the impact of varying parameters $K$ (number of similar queries) and $\alpha$ (threshold score) in the score-based router $R_{K}^{\alpha}$ on execution accuracy (EX) and model distribution. This examination highlights the cost-accuracy trade-off. 
\clearpage
\begin{figure}[t!]
    \centering
    \begin{subfigure}[t]{0.48\textwidth}
        \centering
        \includegraphics[width=\textwidth]{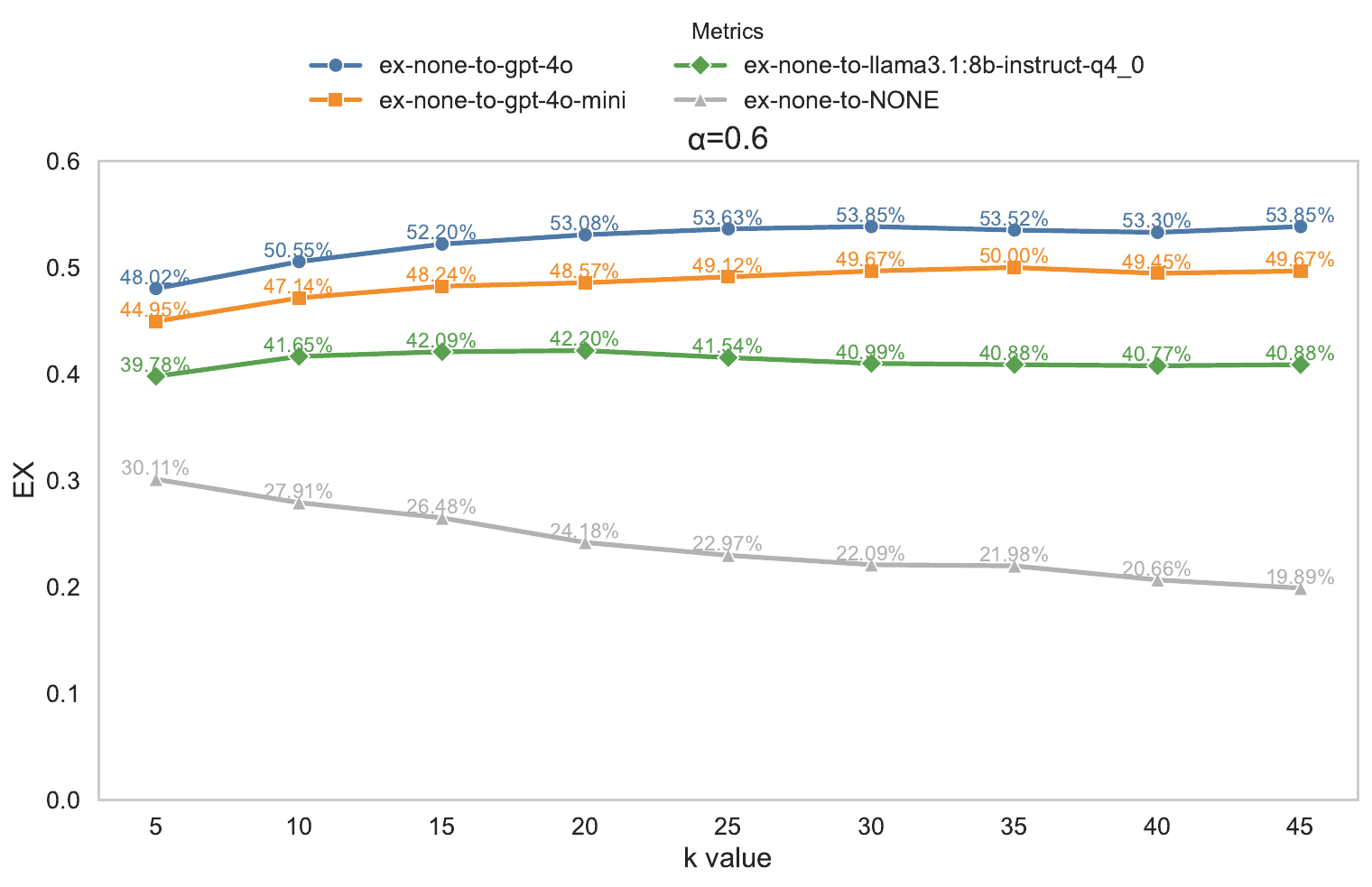}
        \caption{EX as $K$ changes}
        \label{fig:ex_vs_k}
    \end{subfigure}
    \begin{subfigure}[t]{0.48\textwidth}
        \centering
        \includegraphics[width=\textwidth]{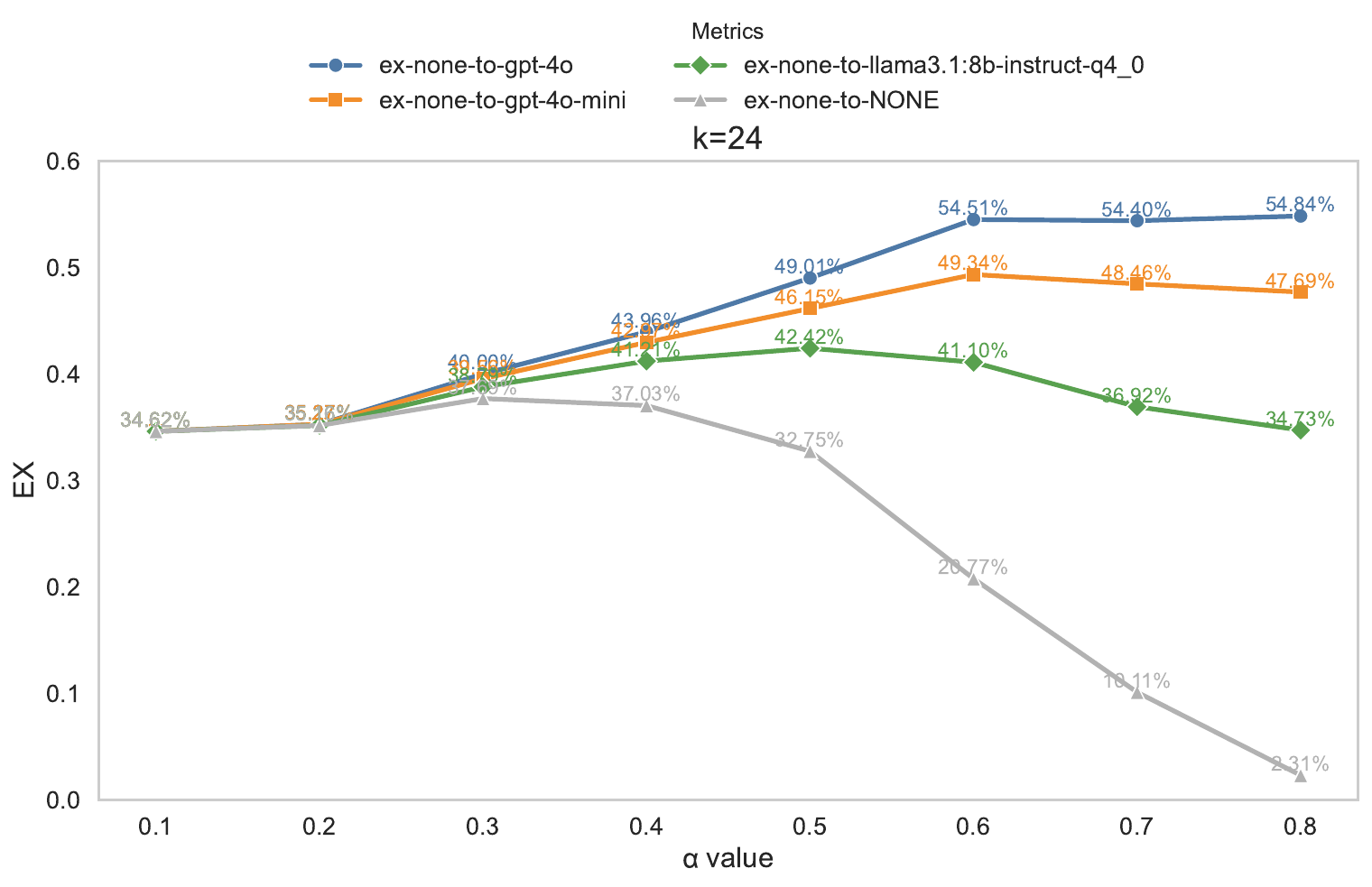}
        \caption{EX as $\alpha$ changes}
        \label{fig:ex_vs_alpha}
    \end{subfigure}
    \caption{Execution Accuracy for a score-based router with different \texttt{None} routing strategies.}
    \label{fig:combined_effects-3}
\end{figure}

\begin{figure}[t!]
    \begin{subfigure}[t]{0.48\textwidth}        
        \includegraphics[width=\textwidth]{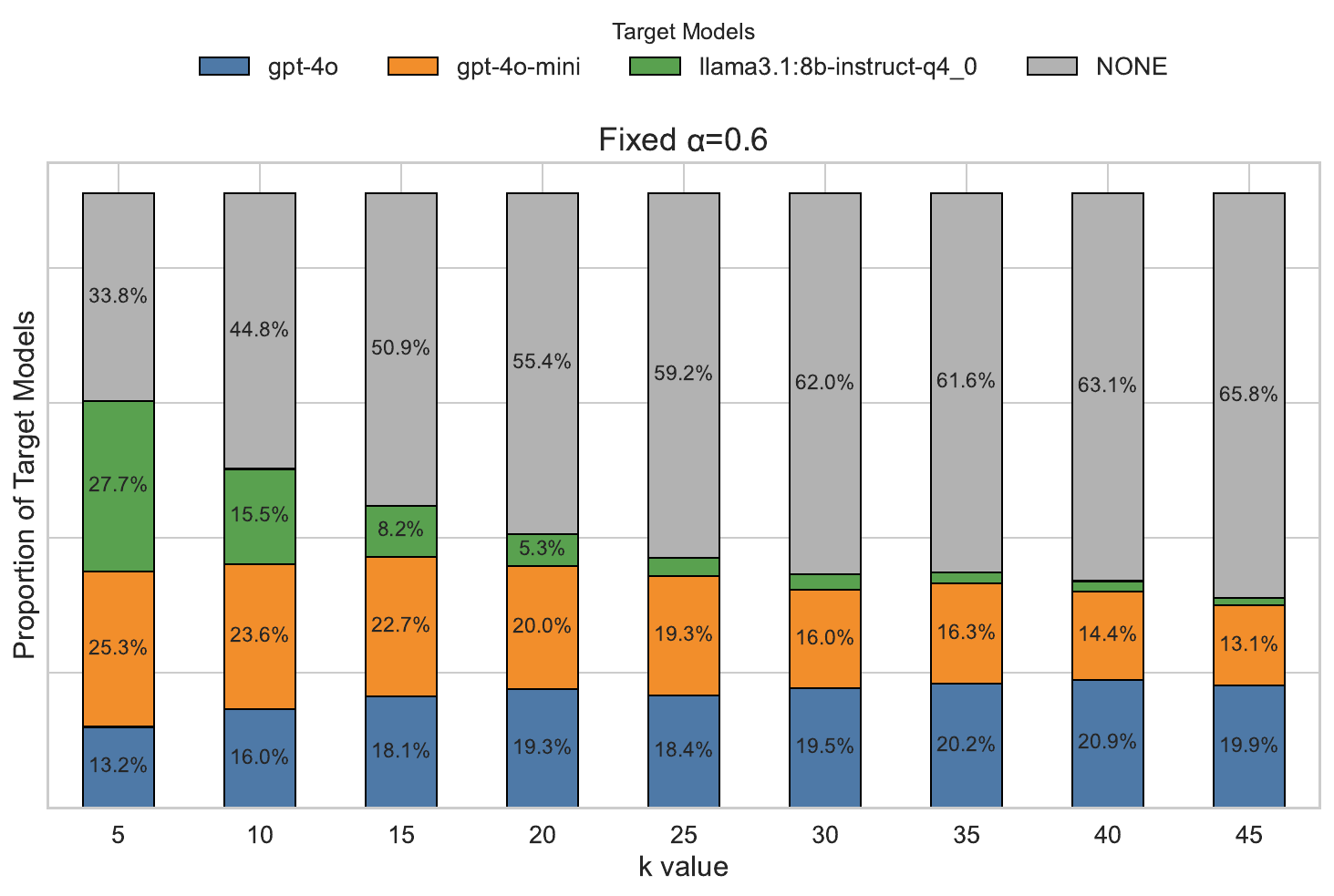}
        \caption{With None prediction}
        \label{fig:distribution_vs_k}
    \end{subfigure}  
    \begin{subfigure}[t]{0.48\textwidth}
        \includegraphics[width=\textwidth]{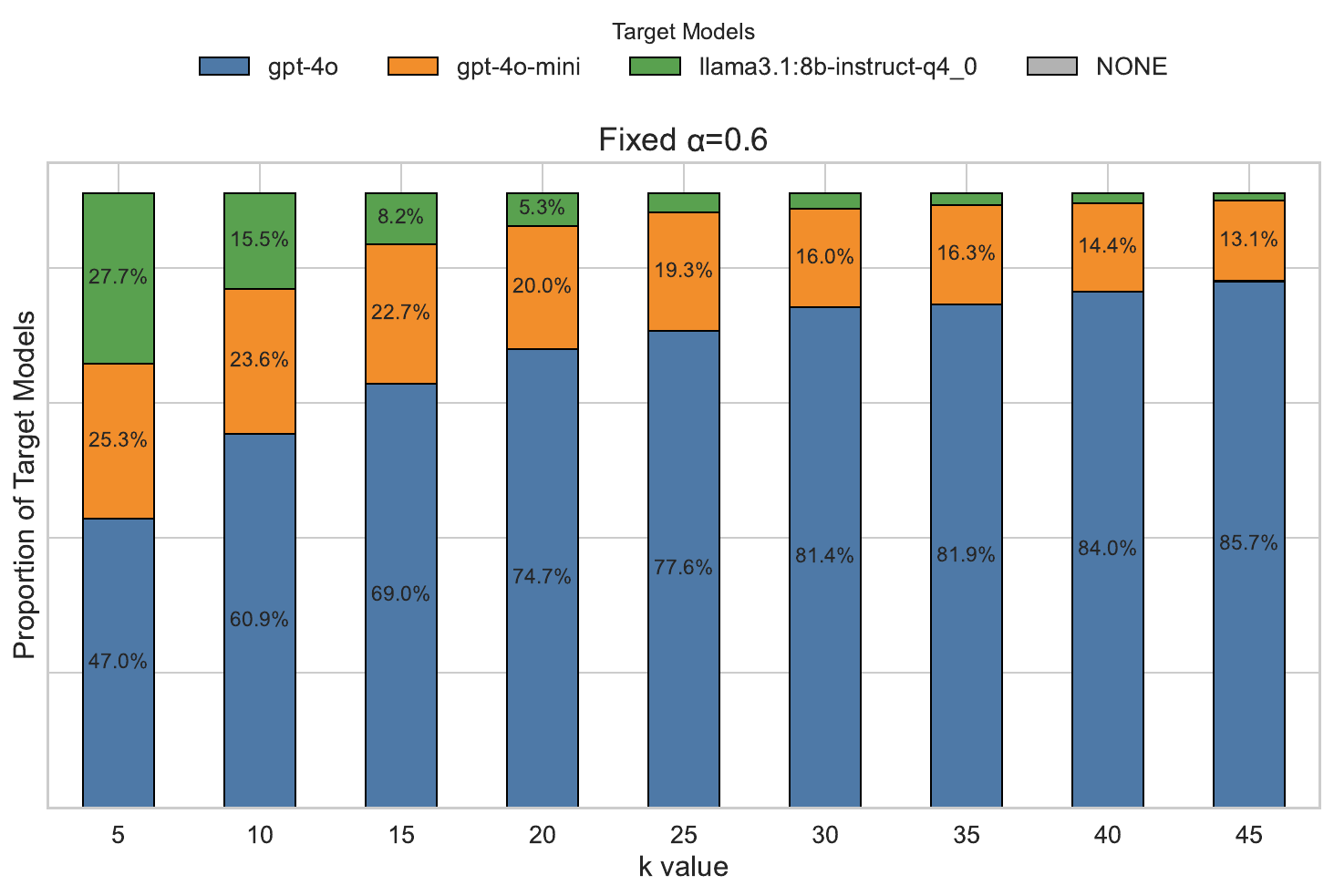}
        \caption{\texttt{None} routed to \texttt{GPT-4o}}
        \label{fig:distribution_vs_k_gpt4o}
    \end{subfigure}
    \caption{Effect of varying $K$ on execution accuracy and on model distribution for score-based router.}
    \label{fig:combined_effects-2}
\end{figure}

\begin{figure}[t!]
    \begin{subfigure}[t]{0.48\textwidth}        
        \includegraphics[width=\textwidth]{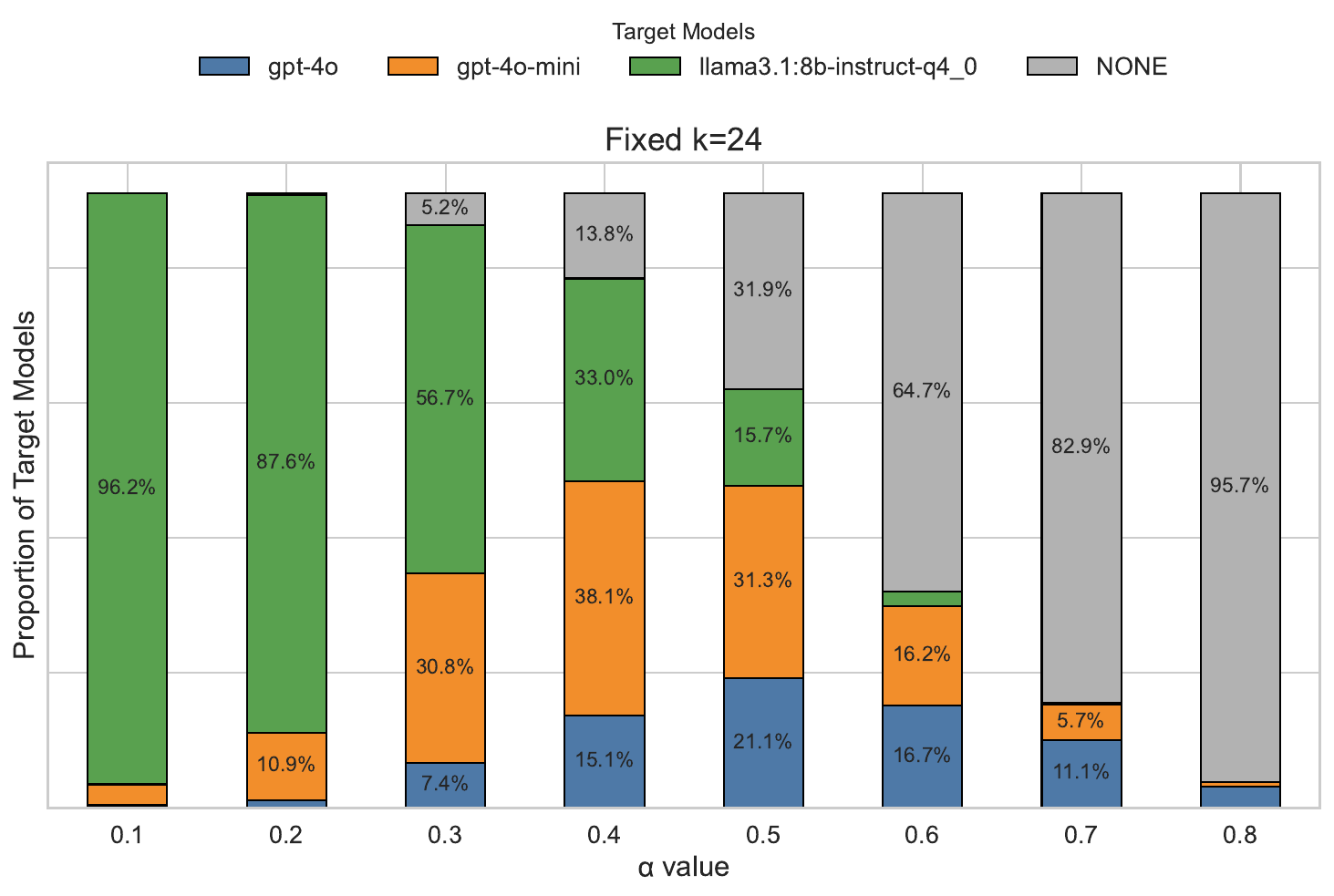}
        \caption{With None prediction}
        \label{fig:distribution_vs_alpha}
    \end{subfigure}
    \begin{subfigure}[t]{0.48\textwidth}        
        \includegraphics[width=\textwidth]{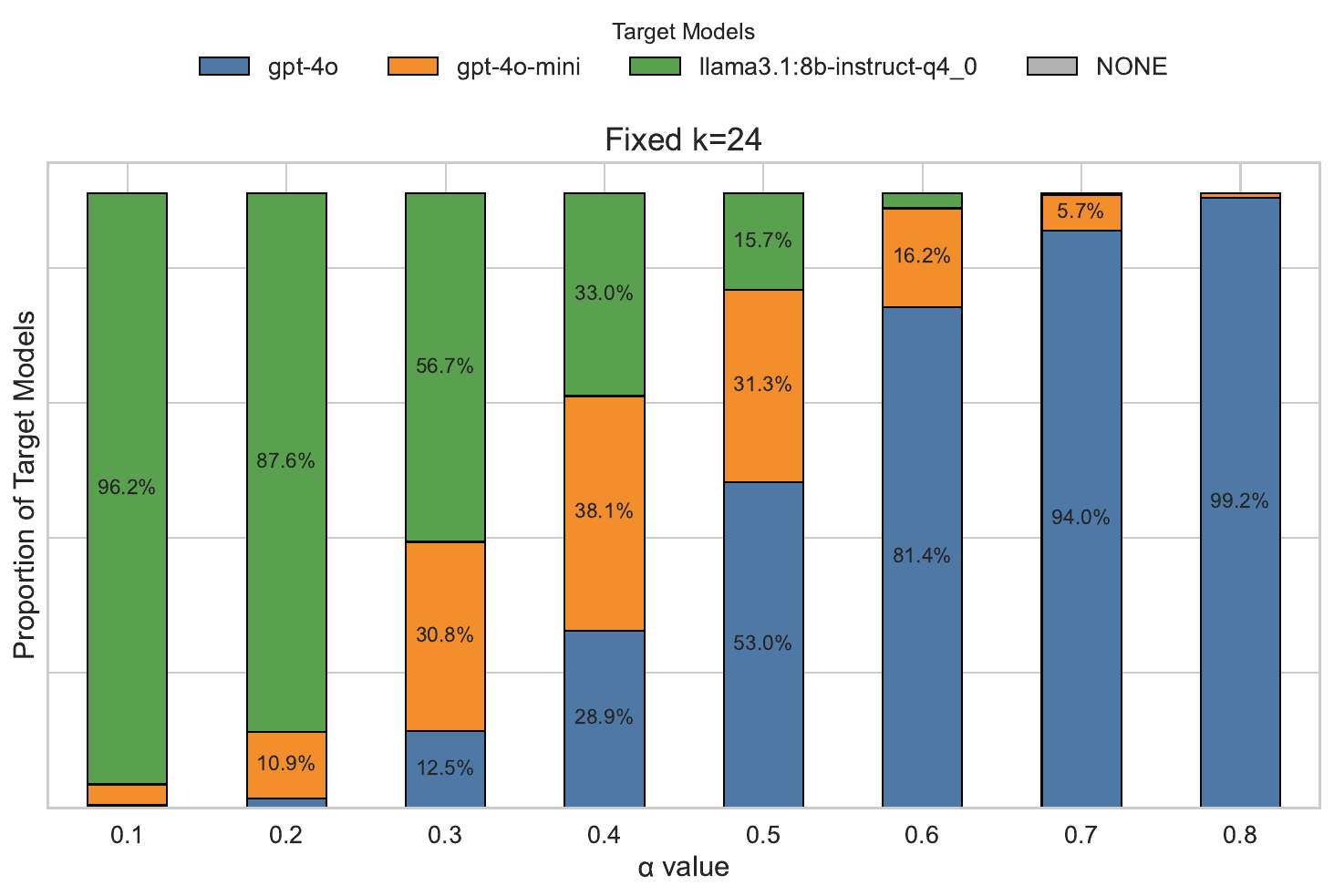}
        \caption{\texttt{None} routed to \texttt{GPT-4o}}
        \label{fig:distribution_vs_alpha_gpt4o}
    \end{subfigure}
    \caption{Effect of varying $\alpha$ on execution accuracy and on model distribution for score-based router.}
    \label{fig:combined_effects}
    \vspace{-1em}
\end{figure}

Our empirical observations are based on 10\% of BIRD's train set, summarized in Figures~\ref{fig:combined_effects-3},~\ref{fig:combined_effects-2}, and ~\ref{fig:combined_effects}:\vspace{-0.5em}

\begin{itemize}
    \setlength{\itemsep}{0pt}
	\setlength{\parskip}{0pt}
	\setlength{\parsep}{0pt}
    \item \textbf{Higher $K$ or $\alpha$ Values:} leads to an increase execution accuracy (EX) but results in a higher proportion of queries being routed to the strongest mode \gpto, increasing overall cost. 
    \item \textbf{Lower $K$ or $\alpha$ Values:} Favor routing to cheaper models like \gptm and \llama, reducing cost but potentially decreasing EX due to their weaker performance.
\end{itemize}

In practical applications, selecting appropriate values for $K$ and $\alpha$ is crucial to balance the need for high execution accuracy with cost efficiency. By adjusting these parameters, practitioners can tailor this score-based router to meet their specific requirements. 

\end{document}